\documentclass[12pt]{article}
\usepackage{epsfig}

\newcommand{\be}{\begin{equation}}
\newcommand{\ee}{\end{equation}}
\newcommand{\bea}{\begin{eqnarray}}
\newcommand{\eea}{\end{eqnarray}}

\date{}
\begin{document}

\title{Element Synthesis in Stars}
\author{F.-K. Thielemann, F. Brachwitz, C. Freiburghaus, E. Kolbe, \\
G. Martinez-Pinedo, T. Rauscher, F. Rembges  \\
Department of Physics \& Astronomy, Univ.~of Basel,\\ 
Klingelbergstrasse 82, CH-4056 Basel, Switzerland\\
\\
W.R. Hix, M. Liebend\"orfer, A. Mezzacappa\\
Physics Division, Oak Ridge National Laboratory,\\ 
Oak Ridge, TN 37831-6371, USA
\\
\\
K.-L. Kratz, B. Pfeiffer
\\
Institut f\"ur Kernchemie, Univ. Mainz, Fritz-Strassmann-Weg 2,\\
D-55099 Mainz, Germany
\\
\\
K. Langanke
\\
Institute of Physics and Astronomy, University of Aarhus,\\
DK-8000 Aarhus C, Denmark
\\
\\
K. Nomoto
\\
Department of Astronomy, University of Tokyo,\\
Tokyo 113-0033, Japan
\\
\\
S. Rosswog
\\
Department of Physics and Astronomy, University of Leicester,\\
University Road, LE1 7RH, Leicester, UK
\\
\\
H. Schatz
\\
National Superconducting Cyclotron Laboratory \& Department of Physics 
\\and Astronomy, Michigan State University, East Lansing, MI 48824
\\
\\
M. Wiescher
\\
Department of Physics, University of Notre Dame, \\Notre Dame, IN 46556, USA
}
%       \thanks{supported by Swiss NSF grants 2000-53798.98 and 2000-061822.00}}
%\input epsfig.sty

\maketitle

\begin{abstract}
Except for $^1$H, $^2$H, $^3$He, $^4$He, and $^7$Li, originating from the
Big Bang, all heavier elements are made in stellar evolution and stellar
explosions. Nuclear physics, and in many cases nuclear structure far from 
stability, enters in a crucial way. Therefore, we examine in this review
the role of nuclear physics in astrophysics in general and in particular how
it affects stellar events and the resulting nucleosynthesis.
Stellar modeling addresses four major aspects: 
1. energy generation and nucleosynthesis, 2. energy
transport via conduction, radiation or possibly convection, 
3. hydrodynamics/hydrostatics, and finally
4. thermodynamic properties of the matter involved. 
Nuclear Physics enters via nuclear reaction cross sections and nuclear
structure (affecting the composition changes and nuclear energy generation),
neutrino-nucleon and neutrino-nucleus cross sections (affecting neutrino
opacities and transport), and e.g.\ the equation of state at and beyond
nuclear densities which creates a relation between the nuclear many
body problem and and hydrodynamic response like pressure and entropy. 
In the following we review
these four topics by highlighting the role and impact of
nuclear physics in each of these aspects of stellar modeling.
The main emphasis is put on the connection to element synthesis.
\end{abstract}

\section{Introduction}

The most dominant influence of nuclear physics on element synthesis is
given by nuclear reactions occurring in a stellar environment. 
Different stellar environments also require different kinds of nuclear physics
knowledge. In stellar evolution all types of fusion reactions 
transform hydrogen to iron-group nuclei via hydrogen, helium, 
carbon, neon, oxygen and silicion burning, releasing eventually
8.7 MeV of binding energy per nucleon. The composition changes in stellar
evolution, occurring on timescales of millions to billions of years rather than
timescales of the strong interaction are dominated by reactions which can barely
proceed for the temperatures involved. The latter  correspond to typical 
bombarding energies below the Coulomb barrier. The study of such reactions 
needs high intensity, (very) low energy beams and either passive or active 
shielding \cite{kaeppeler98,rolfs99}. For more details and updates
see the contribution by C.Rolfs~\cite{rolfs00}.

Reactions occuring in explosive environments experience higher temperatures
and thus higher bombarding energies and larger cross sections. The fact that 
shorter explosive timescales permit reactions for unstable nuclei before their 
decay makes it necessary to investigate also cross sections with unstable
nuclei. For experimental methods and reviews see e.g.\
Refs.\ \cite{kaeppeler98,angulo99}.
Theoretical predictions involving nuclei at excitations with
a sufficiently high density of states, i.e.\ permitting the application of
the statistical model, are reviewed in \cite{rauscher99,rauscher00}, with
a special attention of including isospin mixing, level densities and alpha
potentials. The state of the art in weak interaction predictions,
especially for electron capture on nuclei and the application of the
shell model for large model spaces is discussed in~\cite{langanke00}. 
Many explosive environments with fuels of N$\approx$Z produce
also unstable nuclei close to stability. However, explosive
environments with a large surplus of hydrogen (protons) permit proton
captures up to the proton dripline. 
This gives rise to the so-called rp-process (rapid proton capture). 
Early ignition stages lead to break-out
reactions from reaction cycles similar to the hot CNO~\cite{rembges97a},
so-called two-proton capture reactions via an intermediate proton-unstable
nucleus can play an essential role~\cite{schatz98a}.
Very neutron-rich environments permit
neutron-captures up to the neutron dripline and are the sites of the
r-process (rapid neutron capture). Nuclear structure far from stability enters
directly~\cite{kratz00}, but also features like pygmy resonances in the
E1-strength at low energies (due to extended neutron skins) can have a 
decisive influence on the size of neutron capture cross 
sections~\cite{goriely98}.

Composition changes in astrophysical environments at
low and intermediate temperatures are described by individual cross sections to
follow the reaction flows. High temperatures lead to chemical equilibria.
In the case of the rp- and r-process this causes abundance distributions
on isotonic or isotopic lines which show maxima at specific proton or
neutron separation energies. Therefore, in such cases individual reactions
might not be that important, but such equilibria depend on correct reaction
Q-values or separation energies, i.e.\ the proper knowledge of nuclear
masses far from stability. 
The behavior of shell effects and shell closures deserves special attention 
\cite{kratz00}, see also the contribution by W. Nazarewicz~\cite{nazarewicz00}.
The tendency in recent years is here to move from 
macroscopic-microscopic models like the droplet model to non-relativistic or 
relativistic mean field methods and even shell model calculations,
if the model space permits this 
\cite{moeller95,dobaczewski99,lalazissis99,martinez99}.
Another aspect in the r-process
is related to fission barriers and fission yields for nuclei far from
stability~\cite{cowan91,pearson99,panov01}. 

While strong interaction timescales can support equilibria at high
temperatures (and densities), weak interactions lead very seldom to
chemical equilibria. Exceptions are early phases of the big bang and
equilibrated (cooled) neutron stars. Thus, it is always necessary to follow
each individual weak interaction, i.e.\ beta-decays, electron captures and
neutrino-nucleus interactions. The present state of the art calculations
for beta-decays are QRPA or shell model calculations 
\cite{moeller97,martinez99}, electron capture
calculations are now possible within the shell model up to the
pf-shell, i.e.\ the important Fe-group nuclei \cite{langanke00}. The best 
calculations for
neutrino-nucleus cross sections are available within the continuum RPA
(CRPA) model \cite{kolbe00,hektor00,langanke00a,kolbe01}.

Most of the modes responsible for energy transport in stellar environments 
are not nuclear physics related: mixing/convection of matter with
a given heat content or transport via radiation (photons). The first category
involves all types of hydrodynamic instabilities from convection due to
entropy inversion in regular stellar evolution \cite{heger99} to 
Rayleigh-Taylor instabilities in dynamic events \cite{niemeyer99a,niemeyer99b}
and rotationally induced meridional circulation \cite{maeder98}.
Radiation transport is governed by photon scattering and reaction
processes. Photon-ion interactions with bound-bound and bound-free
transitions are covered in atomic physics and
generally addressed as opacities. In a similar way photon-electron
interactions enter \cite{rogers98}. 
Nuclear physics is addressed when "radiation" transport
proceeds by the way of neutrinos in hot neutron star evironments, be
it in stellar collapse in supernovae or events which involve hot neutron
stars collapsing to black holes (either in massive stars or in neutron
star mergers) \cite{mezzacappa99}. Here the cross sections of neutrino-nucleus, 
neutrino-nucleon,
neutrino-nuclear matter and neutrino-electron/positron collisions are
relevant and the quest is to perform precise calculations
\cite{bruenn91,burrows99,pons99,yamadas00}.

Hydrodynamics is an essential feature in any stellar modeling, be it in 1D
(mostly spherically symmetric)  \cite{mezzacappa00} or 2D and 3D 
\cite{jankmue96,niemeyer99a,niemeyer99b}, multigrid or adaptive grid methods
\cite{dorfi99}, making use of implicit vs.\ explicit numerical
methods \cite{hixthi99}, or employing general relativity rather than 
Newtonian hydrodynamics \cite{liebend2000,liebend01}. 
For these issues, however, there exists no direct
relation to nuclear physics properties, unless one consideres how timescales
of nuclear energy release can enter hydrodynamic modeling.
Nevertheless, hydrodynamics provides the environment conditions which determine
nuclear reactions, composition changes and thus also energy generation.

Convection and radiation transport are both
closely linked to the numerical treatment of hydrodynamics. How
are convective flows modeled in 1D vs.\ multi-D hydro? Which methods are
employed to perform radiation transport, independent on the micro-physics
related cross sections involved? Two of the schemes employed in neutrino
transport are flux-limited diffusion or more recently full transport via
solving the Boltzmann transport equation \cite{messer98,mezzacappa99}.

Thermodynamics is an aspect also coupled strongly to hydrodynamics,
as the pressure due to the equation of state enters directly into the
formation of shock waves, the entropy enters into the formation of
hydrodynamical instabilities and convection. In the outer zones of a
star, especially the atmosphere, the EOS is mainly given by atomic
physics via the mixture of a partially ionized ion, electron, and photon gas
\cite{dappen00}.
In deeper layers ions are fully ionized and the thermalized Fermi gases
(e.g.\ electron/positron), Bose gases of massless particles
(i.e.\ Planck distributions of photons) and Boltzmann distributions of
nuclei are needed. Here nuclear physics enters only indirectly by possibly
governing the reactions which provide the composition of nuclei and
the total amount of electrons available, the latter being dependent on
electron captures and weak interactions in general.

A more direct involvement of nuclear physics is given for the EOS at and
beyond nuclear densities $\rho_0$
\cite{prakash97,latswest91,shen98}. Important 
features are the different stages
of dissolving nuclei into a nucleon soup (involving a number of topologies)
and the creation of new particles at supranuclear 
densities. When do hyperons occur (sigmas, lambdas at $2\times \rho_0$?),
do we have a formation of kaon or pion condensates, at what density
does the quark-hadron phase transition occur (4--5$\times \rho_0$?)
\cite{weber99}?

After this short general introduction and overview, we want to discuss a
number of specific astrophysical applications to show how the different
nuclear aspects affect the synthesis of elements in the modeling of 
stellar events.
In particular we address stellar evolution, type II and type Ia supernovae
(SNe II and SNe Ia),  novae and X-ray bursts (the sites of explosive
H-burning and the rp-process), and the features and possible sites of the 
r-process.

\section{Stellar Evolution}

\subsection{The Role of Individual Reactions}

H-burning converts of $^1$H into $^4$He via pp-chains or the CNO-cycles. The 
simplest PPI chain is initiated by $^1$H($p,e^+\nu$)$^2$H(p,$\gamma$)$^3$He and
completed by $^3$He($^3$He,2p)$^4$He. The dominant CNO-I cycle chain\\
$^{12}$C(p,$\gamma)^{13}$N($e^+\nu)^{13}$C(p,$\gamma)^{14}$N(p,$\gamma)^{15}
$O($e^+\nu)^{15}$N(p,$\alpha)^{12}$C is controlled by the
slowest reaction $^{14}$N($p,\gamma)^{15}$O. Further burning stages are
characterized by their major reactions, which are in 
He-burning $^4$He(2$\alpha,\gamma$)$^{12}$C (triple-alpha) and 
$^{12}$C($\alpha, \gamma$)$^{16}$O, in
C-burning $^{12}$C($^{12}$C, $\alpha$)$^{20}$Ne, and in
O-burning $^{16}$O($^{16}$O,$\alpha$)$^{28}$Si.
The alternative to fusion reactions are photodisintegrations which start to 
play a role at sufficiently high temperatures when 30$kT$$\approx$$Q$ 
(the Q-value 
or energy release of the inverse capture reaction). This ensures the existence
of photons with energies $>$$Q$ in the Planck distribution and 
leads to Ne-Burning [$^{20}$Ne($\gamma,\alpha)^{16}$O, 
$^{20}$Ne($\alpha,\gamma)^{24}$Mg] at $T$$>$$1.5\times 10^9$K (preceding
O-burning) due to a small Q-value of $\approx$4~MeV and Si-burning 
at temperatures in excess of 3$\times$10$^9$K 
(initiated like Ne-burning by photodisintegrations of $^{28}$Si). 
The latter ends in a chemical equilibrium with an abundance 
distribution around Fe (nuclear statistical equilibrium, NSE), as typical 
Q-values of 8--10~MeV permit photodisintegrations as well as
the penetration of the corresponding Coulomb barriers
at these temperatures.
Stars with masses $M$$>$8M$_\odot$ 
develop an onion-like composition structure, after passing through
all hydrostatic burning stages, and produce a collapsing core at the
end of their evolution, which proceeds to nuclear densities 
\cite{nomohash88,woosweav95,chieffi98,heger99,umeda00,heger00}.
Less massive stars experience core H- and He-burning and end as C/O
white dwarfs after strong episodes of mass loss \cite{hashi93}.
These do not exceed the Chandrasekhar mass, i.e.\ the maximum 
stellar mass stabilized against contraction due to the pressure of the
degenerate electron gas.

The major uncertainties in all these hydrostatic burning stages are related
to low energy fusion reactions, i.e.\ the cross sections at sub Coulomb
barrier energies, where (until recently) a determination could be only
obtained by extrapolation to low energies. The status of reactions has been
discussed extensively \cite{kaeppeler98,adelberger98} and is presented in
the most recent compilation \cite{angulo99}. The main breakthrough in recent
years is due to the possibility of measuring cross sections at stellar
thermal energies in underground laboratories (LUNA) \cite{rolfs99,rolfs00}.
A number of uncertain reactions at low energies still wait to be explored
with the aid of this new and promising method.

\subsection{Weak Interactions and Electron Captures}

The late phases of stellar evolution (O- and Si-burning) are less prone
to cross section uncertainties from strong interactions due to the 
emergence of equilibria, as discussed above. However, the high densities
result in partially or fully degenerate electrons with increasing Fermi
energies \cite{nomohash88}. When these supercede the Q-value thresholds of 
electron capture reactions, this allows for electron capture on an increasing 
number of initially Si-group and later Fe-group (pf-shell) nuclei.
Because sd-shell reactions were well understood in the past \cite{fuller85},
O-burning results are quite safe. The recent progress in pf-shell rates
\cite{langanke00} led to drastic changes in the late phases of Si-burning
\cite{heger00}, thus also setting new conditions for the subsequent Fe-core
collapse after Si-burning, the size of the Fe-core and its
electron fraction $Y_e$$=$$<$$Z/A$$>$ \cite{martinez00}.

\section{Massive Stars and Type II Supernovae}

\subsection{The Explosion Mechanism}

Electron captures on pf-shell nuclei (and later on free protons!) in the central
Fe-core, resulting from preceding Si-burning, trigger a pressure decrease
and the collapse of the Fe-core. The size of the homologously collapsing  core
(where infall velocities are proportional
to the radius), which turns into nuclear matter during bounce
\cite{prakash97}, is dependent on the amount of 
prior electron captures on pf-shell nuclei \cite{martinez00,heger00}.
The total energy released, 2--3$\times 10^{53}$erg, equals the
gravitational binding energy of a neutron star and is thus dependent on the
nuclear equation of state (EOS) \cite{prakash97} which determines the central
density. Fig.~1 shows a sequence of density profiles during collapse, bounce,
and shortly thereafter \cite{mezzacappa00,liebend2000},
performed with the Lattimer \& Swesty EOS \cite{latswest91}.

\begin{figure}
\centerline{\epsfig{file=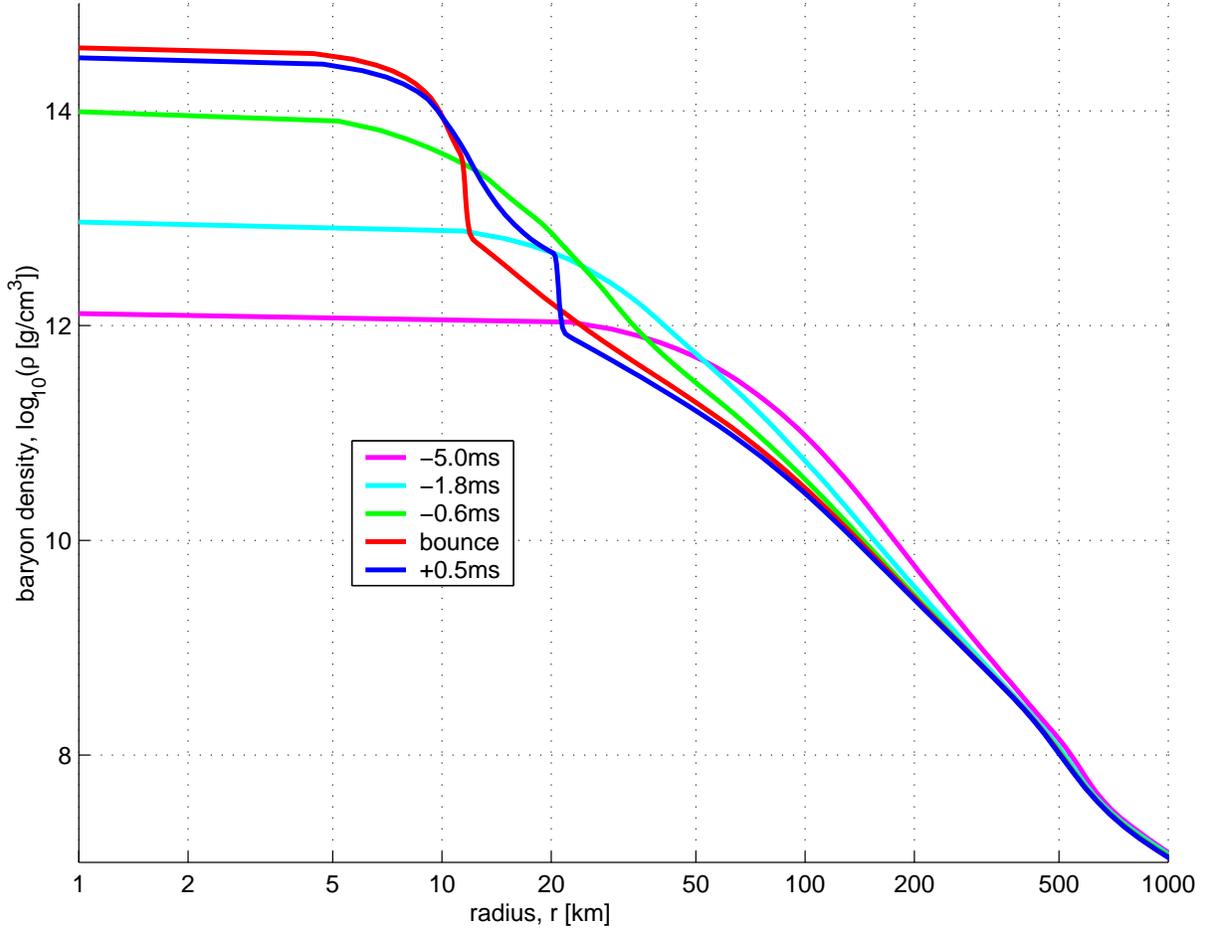,angle=0,width=16cm}}
\caption{A sequence of density profiles of a 13~M$_\odot$ star before and
after core bounce. For such a relatively low mass supernova with a small
Fe-core the bounce occurs at a maximum density of less than twice nuclear 
matter density. At bounce one recognizes the size of the homologous core
(with roughly constant density). After bounce the emergence of an outward
moving density (shock) wave can be witnessed.}
\end{figure}

Because neutrinos are the
particles with the longest mean free path, they are able to carry away that
energy in the most rapid fashion as seen for  
SN1987A in the Kamiokande, IMB and Baksan experiments.
The apparently most promising mechanism for supernova explosions is based on 
neutrino heating beyond the hot proto-neutron star via the dominant processes 
$\nu_e + n \rightarrow p+e^-$ and $\bar\nu_e+p\rightarrow n+e^+$ with a 
(hopefully) about 1\% efficiency in energy deposition 
\cite{jankmue96,mezzacappa99}.
The neutrino heating efficiency depends on the neutrino 
luminosity, which in turn is affected by neutrino opacities 
\cite{bruenn91,burrows99,pons99}, where also inelastic neutrino scattering
on nuclei \cite{kolbe00,langanke00a} contributes. The explosion via 
neutrino heating is delayed after core collapse
for a timescale of seconds or less. Aspects of the explosion mechanism
are still uncertain and depend on Fe-cores from stellar
evolution, electron capture rates of pf-shell nuclei,
the supranuclear equation of state,
as well as the details of neutrino transport \cite{mezzacappa99} and
Newtonian vs.\ general relativistic calculations 
\cite{mezzacappa00,liebend2000,liebend01}.
The observational fact that many core collapse supernovae show polarized
light emission also hints towards a nonspherical
explosion mechanism \cite{khokhlov99}.

\subsection{Composition of Ejecta}

As long as uncertainties are still existing in self-consistent models,
but typical kinetic energies of $10^{51}$~erg are observed in supernova
remnants, light curve as well as explosive nucleosynthesis 
calculations have been performed by introducing a shock of 
appropriate energy in the pre-collapse stellar model 
\cite{woosweav95,thielemann96a,hoffman99,nakamura99,umeda00,rauscher01}.
Such induced calculations, making use of strong and weak interaction nuclear
rates \cite{kaeppeler98,rauscher99,rauscher00,langanke00}, still lack 
self-consistency and cannot predict the
ejected $^{56}$Ni-masses from the innermost
explosive Si-burning layers (powering supernova light curves
by the decay chain $^{56}$Ni-$^{56}$Co-$^{56}$Fe), due to the missing
knowledge of the mass cut between the neutron star and the supernova ejecta.
This relates also to the neutron-richness (or $Y_e$$=$$<$$Z/A$$>$) of the
ejected composition and the weak interactions during stellar evolution
and the explosion.

\begin{figure}
\centerline{\epsfig{file=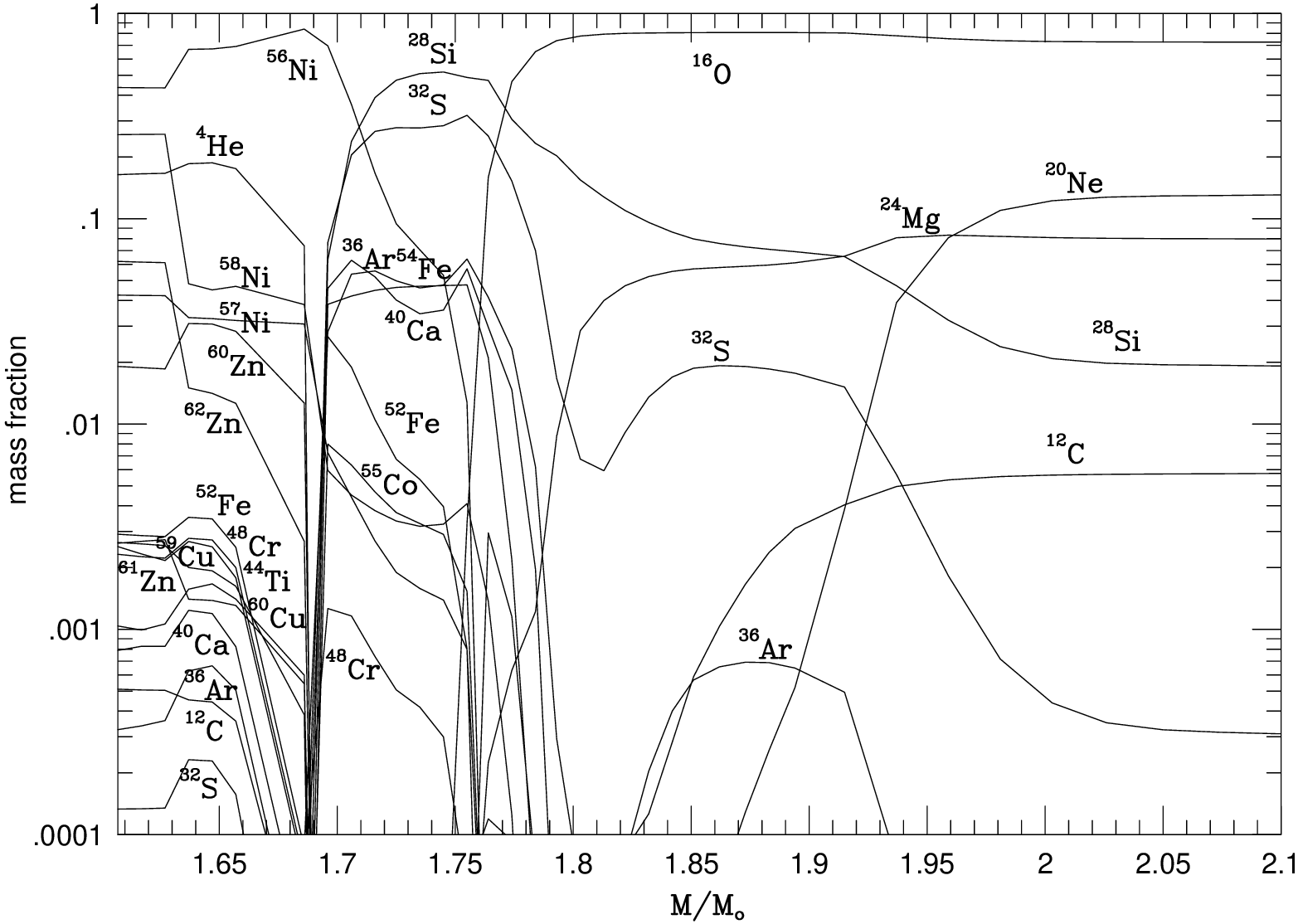,angle=0,width=14cm}}
\vspace{-3cm}
%\centerline{\epsfig{file=thiel.fig1.ps,angle=90,width=10cm}}
\vspace{-2cm}
\caption{Isotopic composition for the inner layers of a core collapse          
supernova from a 20 M$_\odot$ progenitor star with a 6 M$_\odot$   
He-core and a net explosion energy of $10^{51}$ erg, remaining in kinetic
energy of the ejecta. $M(r)$ indicates the radially enclosed mass, integrated
from the stellar center.
The exact mass cut in $M(r)$ between neutron star and 
ejecta depends on the details of the delayed explosion mechanism.
}
\end{figure}
           
Fig.~2 shows the composition after explosive processing 
\cite{hixthi99,thielemann96a} due to the shock wave causing a supernova 
explosion. The outer ejected layers ($M(r)$$>$2M$_\odot$) are unprocessed by 
the explosion and contain results of prior H-, He-, C-, and 
Ne-burning in stellar evolution. 
The interior parts of SNe II contain products of explosive Si, O, and Ne
burning.  In the inner ejecta, which experience 
explosive Si-burning, $Y_e$ changes from 0.4989 to 0.494. 
The $Y_e$ originates from beta-decays and electron captures in the 
pre-explosive hydrostatic fuel and possible alterations via neutrino
reactions during the explosion in these layers.
Huge changes occur in the Fe-group composition for mass zones    
below $M(r)$=1.63M$_\odot$. 
There the abundances of $^{58}$Ni and           
$^{56}$Ni become comparable. All neutron-rich isotopes increase                 
($^{57}$Ni, $^{58}$Ni, $^{59}$Cu, $^{61}$Zn, and $^{62}$Zn), the even-mass 
isotopes ($^{58}$Ni and $^{62}$Zn) show the strongest effect.              
One can also recognize        
the increase of $^{40}$Ca, $^{44}$Ti, $^{48}$Cr, and $^{52}$Fe for the inner
high entropy zones, but a reduction of these $N$=$Z$ nuclei
in the more neutron-rich layers. More details can be found in extended
discussions \cite{thielemann96a,thielemann98,nakamura99}.

Recent calculations \cite{rauscher01,heger00a,rauscher01a} included all nuclides
important for explosive stellar burning (with the exception of a
possible r-process in the neutrino wind above the proto-neutron star,
see Section \ref{pstudy}) in an extended reaction network up to Bi.
Thus, it became possible to follow the $\gamma$-process \cite{woohow78}
(also known as the p-process \cite{rayet90}) self-consistently. Because
of the high temperatures in the explosion previously produced heavy,
stable isotopes can be photodisintegrated (mainly via ($\gamma$,n) and
($\gamma$,$\alpha$) reactions) which leads to the formation of
proton-rich stable isotopes, called the p-isotopes for historical
reasons (indicated as P in Fig.\ 6). Currently, the $\gamma$-process is
considered to be the source of the p-elements, however, the latest
results may also still allow for a (currently hypothetical) additional
contribution from type I X-ray bursts for the low mass p-nuclei (see
Section \ref{sec:rpfinal}).

A correct prediction of
the amount of Fe-group nuclei ejected (which includes also one of the 
so-called alpha elements, i.e.\ Ti) and their relative composition
depends directly on the explosion mechanism and the size of 
the collapsing Fe-core. 
Three types of uncertainties are inherent in the Fe-group ejecta,
related to (i) the total amount of Fe(group) nuclei ejected and the
mass cut between neutron star and ejecta, mostly
measured by $^{56}$Ni decaying to $^{56}$Fe, (ii) the total explosion energy
which influences the entropy of the ejecta and with it the amount of
radioactive $^{44}$Ti as well as $^{48}$Cr, the latter decaying later to
$^{48}$Ti and being responsible for elemental Ti, and (iii) finally
the neutron richness or $Y_e$=$<Z/A>$ of the ejecta,
dependent on stellar structure, electron captures and neutrino interactions.
$Y_e$ influences strongly the ratios of isotopes 57/56 in
Ni(Co,Fe) and the overall elemental Ni/Fe ratio. The latter being dominated
by $^{58}$Ni and $^{56}$Fe. 

The pending understanding of the explosion mechanism also affects possible
r-process yields for SNe II.
Some calculations seemed to be able to reproduce the solar
r-process abundances well in the high entropy neutrino wind, emitted from
the hot protoneutron star after the SN II explosion 
\cite{takahashi94,woosley94,qian96b}.
However, present-day supernova models have 
difficulties to reproduce the entropies required for such abundance 
calculations and in addition face problems with abundance features in the
mass range 80-120 \cite{freiburghaus99a}.
The inclusion of non-standard neutrino properties may 
perhaps achieve low enough $Y_e$'s for intermediate entropies to correct for
such unwanted features \cite{mclaughlin99}.
However, recent observations shed some doubts on the supernova origin.
On average SNe II produce Fe to intermediate mass elements
in ratios within a factor of 3 of solar. 
If they would also be responsible for the r-process, the same limits should 
apply. But the observed bulk r-process/Fe ratios vary
widely in low metallicity stars by more than a factor of 100
\cite{sneden00,truran00}.

\section{Type Ia Supernovae}

\subsection{Thermonuclear Explosions}

      There are strong observational and theoretical indications that
SNe Ia are thermonuclear explosions of accreting white dwarfs in binary
stellar systems \cite{hoefkhok96,nugent97,nomoto00,livio99}. 
High rates of H-accretion cause
high temperatures at the base of the accreted matter and lead to quasi-stable
H-burning and subsequent He-burning in shells surrounding the white dwarf,
possibly related to supersoft X-ray sources.
This increases the mass of the white dwarf consisting of C and O towards
the maximum stable Chandrasekhar mass and leads to
contraction. 

Contraction causes carbon ignition in
the central region and a thermonuclear runway with a complete explosive
disruption of the white dwarf \cite{nomthiyok84,woosweav94}. 
High accretion rates cause a higher central temperature and pressure, 
favoring lower ignition densities.  A flame front
then propagates at a subsonic speed as a deflagration wave due
to heat transport across the front \cite{niemeyer99b,hillnie00}.  
Here the most uncertain quantity is the flame speed which
depends on the development of instabilities of various scales at the
flame front.  Multi-dimensional hydro simulations of the
flame propagation have
suggested that a carbon deflagration wave might propagate at a speed
$v_{\rm def}$ as slow as a few percent of the sound speed $v_{\rm s}$
in the central region of the white dwarf.  
The nucleosynthesis consequences of such slow flame speeds witness the
actual burning front velocities and can thus serve as a constraint.  
Electron capture affects the central electron fraction $Y_e$,
which determines the composition of the ejecta from such explosions.
The amount of electron capture depends on 
(i) the electron capture rates of pf-shell nuclei,
(ii) $v_{\rm def}$,
influencing the time duration of matter at high temperatures (and with
it the availability of free protons for electron captures), and (iii) the 
central
density of the white dwarf $\rho_{ign}$
(increasing the electron chemical potential i.e.\ their Fermi energy)  
\cite{iwamoto99,brachwitz00,langanke00}.

After an initial deflagration in the central layers, the
deflagration can turn into a detonation (supersonic burning front)
at lower densities \cite{niemeyer99a}.
The transition from a deflagration to a detonation (delayed detonation
model) leads to a change in the ratios of Si-burning subcategories
with varying entropies.
This also leaves an imprint on the Fe-group composition.

\subsection{Nucleosynthesis Details}

Nucleosynthesis constraints can help to find the "average" SN Ia
conditions responsible for their contribution to
galactic evolution, i.e.\ especially the Fe-group composition.
SNe Ia contribute essentially no elements lighter than Al, about 1/3 of the
elements from Si to Ca, and the dominant amount of Fe group nuclei (Ti to Ni).
In addition, the average Fe-group yields of SNe II and SNe Ia differ.

\begin{figure}
\centerline{\epsfig{file=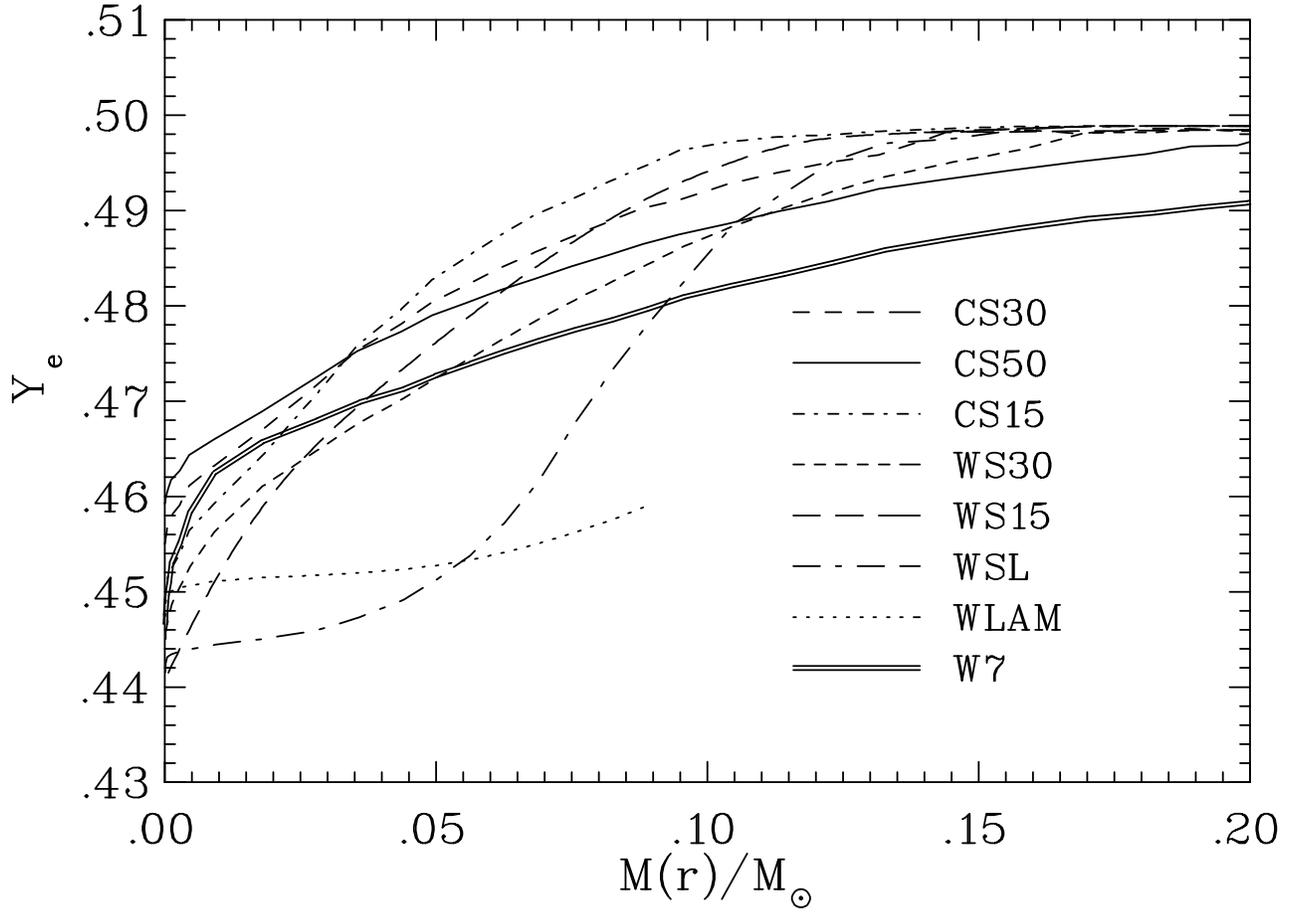,width=12cm,angle=90}}
\caption{$Y_e$ after freeze-out of nuclear reactions 
measures the electron captures on free protons and 
nuclei. Small
burning front velocities lead to steep $Y_e$-gradients which flatten
with increasing velocities (see the series of models CS15, CS30, and CS50
or WS15, WS30, and W7).
Lower central ignition densities shift the curves up (C vs.\ W), but the
gradient is the same for the same propagation speed. Only when
the $Y_e$ from electron captures is smaller
than for stable Fe-group nuclei, subsequent $\beta^-$-decays
will reverse this effect (WSL and WLAM).}
 \end{figure}

Fig.~3 shows the influence of central ignition densities $\rho_{ign}$ 
     1.37 (C) and 2.12$\times 10^9$ g cm$^{-3}$ (W) at the onset of the
thermonuclear runaway and slow (S) deflagration speeds of
$v_{\rm def}/v_{\rm s}$ = 0.015
(WS15, CS15), 0.03 (WS30,CS30) or 0.05 (CS50) on the resulting $Y_e$ due to
the different amount of electron capture \cite{iwamoto99}.
$Y_e$ values of 0.47-0.485 lead to dominant abundances of
$^{54}$Fe and $^{58}$Ni, values between 0.46 and 0.47 produce
dominantly $^{56}$Fe, values in the range of 0.45 and below are
responsible for $^{58}$Fe, $^{54}$Cr, $^{50}$Ti, $^{64}$Ni, and values
below 0.43-0.42 are responsible for $^{48}$Ca.  
The intermediate $Y_e$-values 0.47-0.485 exist in all cases, but the masses
encountered which experience these conditions depend on the $Y_e$-gradient
and thus $v_{def}$. Whether the lower vales with $Y_e$$<$0.45 are attained,
depends on the central ignition density $\rho_{ign}$. Therefore, 
$^{54}$Fe and $^{58}$Ni are indicators of $v_{def}$ while  $^{58}$Fe, 
$^{54}$Cr, $^{50}$Ti, $^{64}$Ni, and $^{48}$Ca are a measure of $\rho_{ign}$.
These are the (hydrodynamic) model parameters. An additional uncertainty
is the central C/O ratio of the exploding white dwarf~\cite{hoeflich01}.
Nuclear uncertainties based on electron capture rates are addressed in 
the following subsection.

\subsection{The Effect of Electron Capture Rates}

As the electron gas in white dwarfs is degenerate, characterized by high
Fermi energies for the high density regions in the center, electron capture on
intermediate mass and Fe-group nuclei plays an important role in explosive
burning. Electron capture affects the central electron fraction $Y_e$,
which determines the composition of the ejecta from such explosions.
Recently new Shell Model Monte Carlo (SMMC) and large-scale
shell model diagonalization calculations have been performed for
pf-shell nuclei \cite{dean98,langanke00}. These lead in 
general to a reduction 
of electron capture rates in comparison with previous, more phenomenological, 
approaches. Making use of these new shell model based rates, we present the 
results from Brachwitz et al. \cite{brachwitz00} for the composition of 
Fe-group nuclei produced in
the central regions of SNe Ia and possible changes in the constraints on model
parameters
like ignition densities $\rho_{ign}$ and burning front speeds $v_{def}$,
superceding the results of Fig.~3 \cite{iwamoto99}. 

For a better understanding of the results we employed four rate sets:
(i) The original FFN rates by Fuller et al. \cite{fuller85} as a 
benchmark for further comparisons; (ii)
inclusion of the electron captures rates calculated
within the SMMC method \cite{dean98}, replacing the corresponding 
FFN rates. SMMC rates were used for the parent nuclei $^{45}$Sc,$^{48,50}$Ti,
$^{51}$V,$^{50,52}$Cr,$^{55}$Mn,$^{54-56,58}$Fe, and $^{55,57,59}$Co,
$^{56,58,60}$Ni, otherwise rates were taken from FFN;
(iii) rates from large-scale shell model diagonalization calculations 
\cite{langanke00}, labeled with SMFA.
(iv) As a further option we treated even-even (ee), odd-A (oa), and odd-odd (oo)
nuclei in different
ways, in order to test the sensitivity of the models and the importance of
the rates of particular nuclei.
Such calculations are denoted by SMFA with the corresponding extension
ee, oa, oo or by combinations, e.g.\ ee+oa.
With these modifications of the electron capture rates,
the nucleosynthesis for the SN Ia models WS15 and CS15 \cite{iwamoto99}
was recalculated \cite{brachwitz00}. (It should be mentioned here that the
SMMC rate set suffers from missing odd-odd nuclei.)

\begin{figure}
\vspace{-3cm}
\epsfig{file=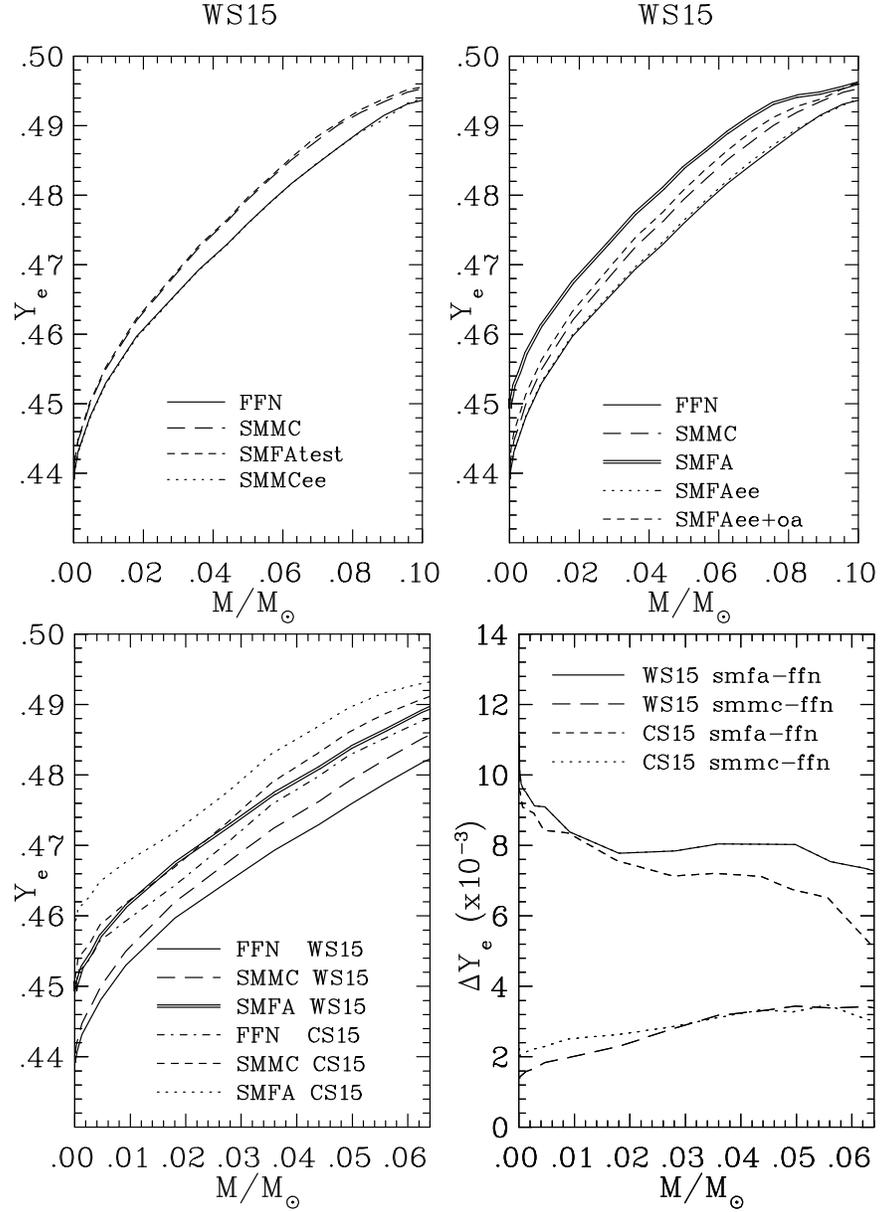,width=16cm,angle=0}
\vspace{-2cm}
\caption{$Y_{e}$, the total proton to nucleon ratio and thus a measure of
electron captures on free protons and nuclei, after freeze-out of nuclear
reactions, as a function of radial mass
for different models and electron capture rates. Also the $Y_{e}$-difference
$\Delta Y_{e}$ between various cases is shown at the bottom right (d).
A detailed discussion of the changes with different electron capture
rate sets is given in the text. 
\label{yemr}}
\end{figure}

The resulting $Y_{e}$-curves (Fig.~\ref{yemr}b) displays a small
$Y_{e}$ shift between SMMC and SMFAee+oa, and a larger
$Y_{e}$-shift between SMFAee+oa and SMFA.
Therefore, the inclusion of odd-odd nuclei has the largest influence on
the $Y_{e}$ difference between SMFA and SMMC.
The rate change for odd-A nuclei is mostly responsible
for the $Y_{e}$-shift between FFN and SMMC, and the inclusion of odd-odd
nuclei causes the largest part of the $Y_{e}$-shift between SMMC and SMFA.
Thus, the changes in the electron capture rates for
odd-A and odd-odd nuclei are responsible
for the $Y_{e}$ difference between SMFA and FFN, while the contribution
of even-even nuclei is negligible, an assertion which was directly tested by
case SMFAee (Fig.~\ref{yemr}b). 

Notice, however, that the changes for
a given model (here WS15 and CS15) lead to almost parallel $Y_e$-curves
in the intermediate $Y_e$-range responsible for the major abundances
of $^{54}$Fe and $^{58}$Ni. This can also be seen in the close to constant
$\Delta Y_e$-curves in Fig.~\ref{yemr}d. Thus, a change in electron capture 
rates does (to first order) not affect the $Y_e$-gradient of a model,
which is therefore only determined by $v_{def}$ \cite{iwamoto99}.
Therefore, we can conclude that the consequences for the permitted 
range of burning front speeds remain the same. 
The conclusions to be drawn from these results are that:
(i) $v_{def}$ in the range 1.5--3\% of the sound speed is preferred
(cases 15 and 30 over 50) \cite{iwamoto99}, and (ii) the change in pf-shell 
electron capture rates \cite{langanke00} made it possible to have
ignition densities as high as 
$\rho_{ign}$$=$$2\times 10^{9}$ g cm$^{-3}$ without distroying the agreement
with solar abundances of very neutron-rich species \cite{iwamoto99,brachwitz00}.
It seems, however, hard to produce amounts of $^{48}$Ca sufficient to
explain solar abundances in the exploding white dwarfs with these changes 
and more realistic C/O ratios~\cite{woosley98,brachwitz01}.

\section{The rp-Process and X-ray Bursts}

\subsection{Explosive hydrogen burning}

The major astrophysical events which involve explosive H-burning are novae
and type I X-ray bursts. Both are related to binary stellar systems with
hydrogen accretion from a binary companion onto a compact object, and the
explosive ignition of the accreted H-layer. High densities
cause the pressure to be dominated by the degenerate electron gas,
preventing a stable and controlled burning. In the case of novae the compact 
object is a white dwarf, in the case of X-ray bursts it is a neutron
star. Explosive H-burning in novae has been discussed in many recent articles
\cite{jose99,starrfield99,starrfield00,coc00}. Its processing is limited
due to maximum temperatures of $\sim$3$\times 10^8$K.
Only in X-ray bursts temperatures larger than 4$\times 10^8$K are possible.

\subsection{Type I X-ray bursts}

In type I X-ray bursts \cite{taam85,lewin93,taam96,schatz98a,wiescher00}
hydrogen (and helium) burn explosively in a thermonuclear runaway.
In essentially all nuclei below Ca, the proton capture reaction on the
nucleus ($Z_{even}$--1,$N$=$Z_{even}$) produces the compound nucleus
above the alpha-particle threshold and permits a $(p,\alpha)$ reaction. 
This is typically not the case for ($Z_{even}$--1,$N$=$Z_{even}$--1) due to 
the smaller proton separation energy, which leads to hot CNO-type cycles
above Ne~\cite{thielemann94a}. There is one exception, $Z_{even}$=10, where the 
reaction $^{18}$F($p,\alpha)$ is possible, avoiding $^{19}$F
and a possible leak via $^{19}$F($p,\gamma)$ into the NeNaMg-cycle.
This has the effect that only alpha induced reactions like $^{15}$O($\alpha,
\gamma)$ can aid a break-out from the hot CNO-cycle
to heavier nuclei beyond Ne~\cite{wiescher99}.
Break-out reactions from the hot CNO-type cycles above Ne proceed typically via
proton captures on the nucleus ($Z_{even}$,$N$=$Z_{even}$-1)
\cite{thielemann94a,rembges97a}. They occur
at temperatures of about $3\times 10^8$K, while the alpha-induced break-out 
from the hot CNO-cycle itself is delayed to about $4\times 10^8$K.
\begin{figure}[tbp]
\epsfig{file=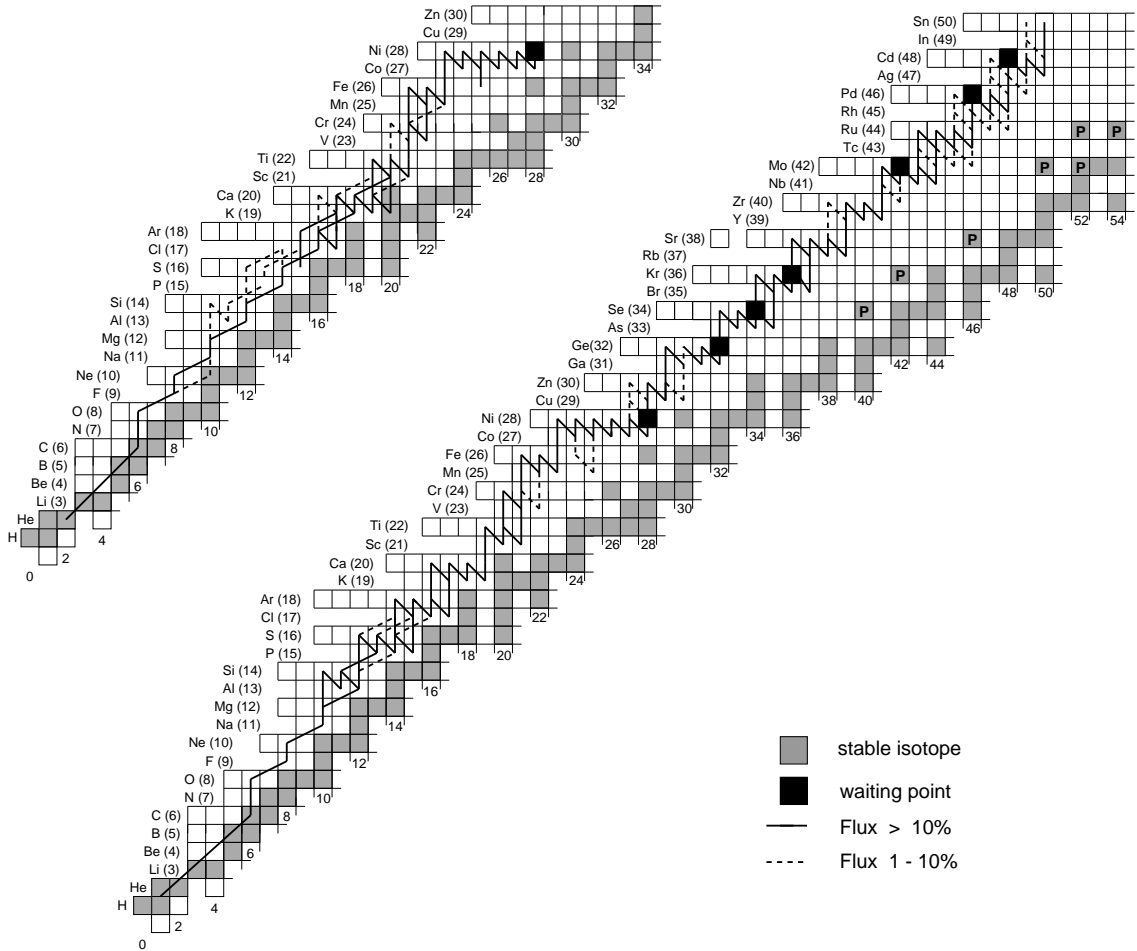,width=15cm,angle=0}
\caption[F1]{
rp and $\alpha$p-process flow up to and beyond Ni. The reaction
flows shown in the nuclear chart are integrated reaction fluxes from a 
time dependent network calculation \cite{schatz98a}, (a) during the
initial burst and thermal runaway phase of about 10s, (b) after the onset of 
the cooling phase when the proton capture on $^{56}$Ni is not blocked anymore 
by photodisintegrations (extending for about 200s). Waiting points above 
$^{56}$Ni are represented by filled squares, stable nuclei by hatched squares,
light p-process nuclei below A=100 are indicated by a $P$.}
\label{rp1}
\end{figure}

In the next stage of the ignition process also He is burned
via the 3$\alpha$-reaction and the $\alpha$p-process (a sequence
of ($\alpha,p$) and ($p,\gamma$) reactions), which 
provides seed nuclei for hydrogen burning via the rp-process
(proton captures and beta-decays).
Processing of the rp-process beyond $^{56}$Ni is 
shown in  Fig.~\ref{rp1} \cite{schatz98a}.
Certain nuclei play the role of long waiting points in 
the reaction flux, where long beta-decay half-lives
dominate the flow, either competing with slow ($\alpha,p$) reactions
or negligible $(p,\gamma)$ reactions, because they are inhibited by inverse
photodisintegrations for the given temperatures. Such nuclei were identified 
as $^{25}$Si ($\tau_{1/2}$=0.22s), $^{29}$S (0.187s),
$^{34}$Ar (0.844s), $^{38}$Ca (0.439s) \cite{wiescher00}. 
The bottle neck at $^{56}$Ni
can only be bridged for minimum temperatures around $10^9$K (in order to
overcome the proton capture Coulomb barrier), maximum temperatures below
$2\times 10^9$K (in order to avoid photodisintegrations), and high densities 
exceeding $10^6$g~cm$^{-3}$ \cite{schatz98a,rehm98}.
If this bottle neck can be overcome, other waiting points like $^{64}$Ge
(64s), $^{68}$Se (96s), $^{74}$Kr (17s) seem to be hard to pass. However,
partially temperature dependent half-lives (due to excited state population),
or mostly 2p-capture reactions (introduced in \cite{goerres95} and
applied in \cite{schatz98a}) can help.

\subsection{The final composition}
\label{sec:rpfinal}

The initial break-out reactions from hot CNO-type cycles and the
hold-ups at waiting points introduce a time structure in energy generation. 
One of the essential questions is whether they can/will show up in 
the X-ray light-curves of bursts. The other question is whether 
individual proton-capture reactions matter, because at peak temperatures 
$(p,\gamma)$-$(\gamma,p)$ equilibria are attained, leading to an equilibrium
distribution along isotonic lines, where only beta-decays might matter.
The latter was claimed~\cite{iliadis99} based on an X-ray burst temperature
profile provided by F. Rembges~\cite{rembges99}. This paper showed (as should 
be expected from the late equilibrium conditions) that with a given temporal 
temperature profile the resulting composition does not depend on individual 
reactions. Thus, the important question is whether the feedback from
hydrodynamics, due to the changed energy input in the early ignition stages
when the break-out reactions occur, leads to different temperature profiles
and thus different final abundances. 

\begin{figure}[tbp]
\epsfig{file=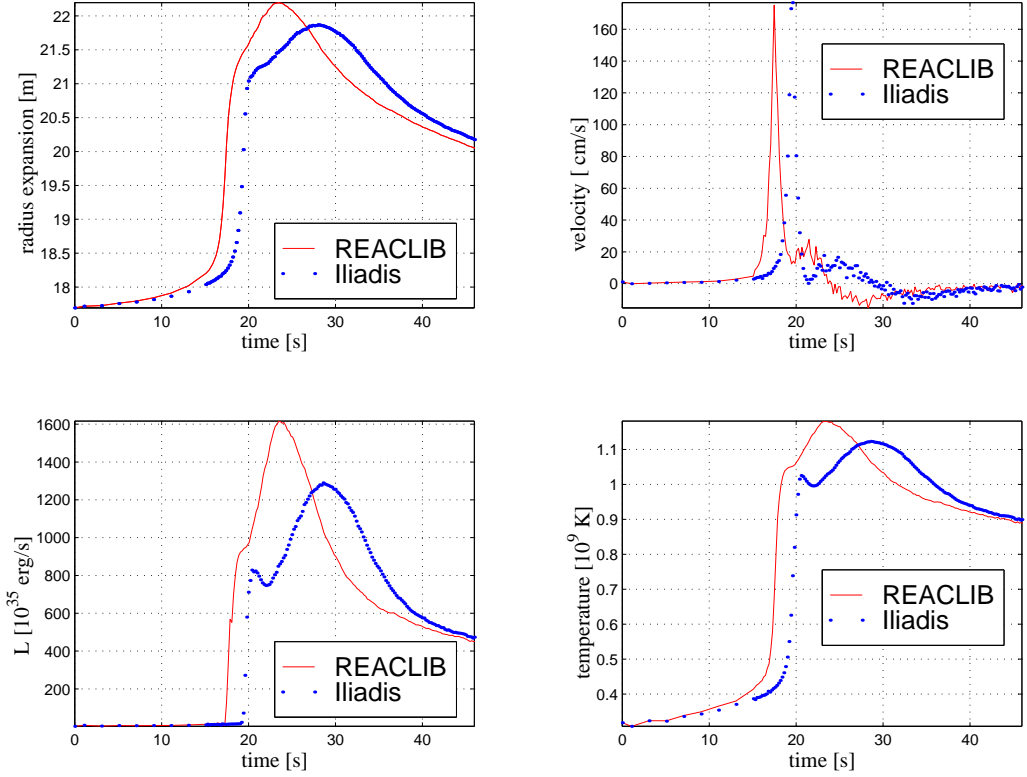,width=16cm,angle=0}
\caption[F1]{
Comparison of burst profiles with different proton capture rates for
the break-out reactions on $^{27}$Si, $^{31}$S, $^{35}$Ar, and $^{38}$Ca
in a self-consistent X-ray burst model~\cite{rembges99,rembges00}. 
Due to different rates the burning of matter beyond Ne is burned on different
timescales, causing a different pre-expansion before the maximum temperature 
are attained in alpha-induced recations.}
\label{rp2}
\end{figure}
The results of a self-consistent calculation can be seen in Fig.~\ref{rp2}
from \cite{rembges99,rembges00,rauscher00s} 
which shows the radius of the burning shell,
the velocity, the luminosity, and the temperature. The difference between
calculations with the old REACLIB rates~\cite{thielemann87} and the ones
with updated cross sections for the break-out points $^{27}$Si, $^{31}$S,
$^{35}$Ar, and $^{38}$Ca emerges at the expected temperatures of
$(3-4)\times 10^8$K, when these break-out reactions occur. At that point
the nuclei beyond Ne will burn towards Ni/Fe. This energy input causes
the temperature increase which will then permit the hot CNO break-out via
alpha-induced reactions and later on also the triple-alpha link to $^{12}$C.
This leads to the burning of $^4$He and the temperature peak for the
rp-process with a chemical equilibrium for proton-capture reactions.
However, as the initial break-out phase differs, different pre-expansions
occur, causing different densities and also different peak temperatures.

\begin{figure}[th]
%\plotone{/d5/fkt/sanibel/tiac_1q.ps}
\centerline{\epsfig{file=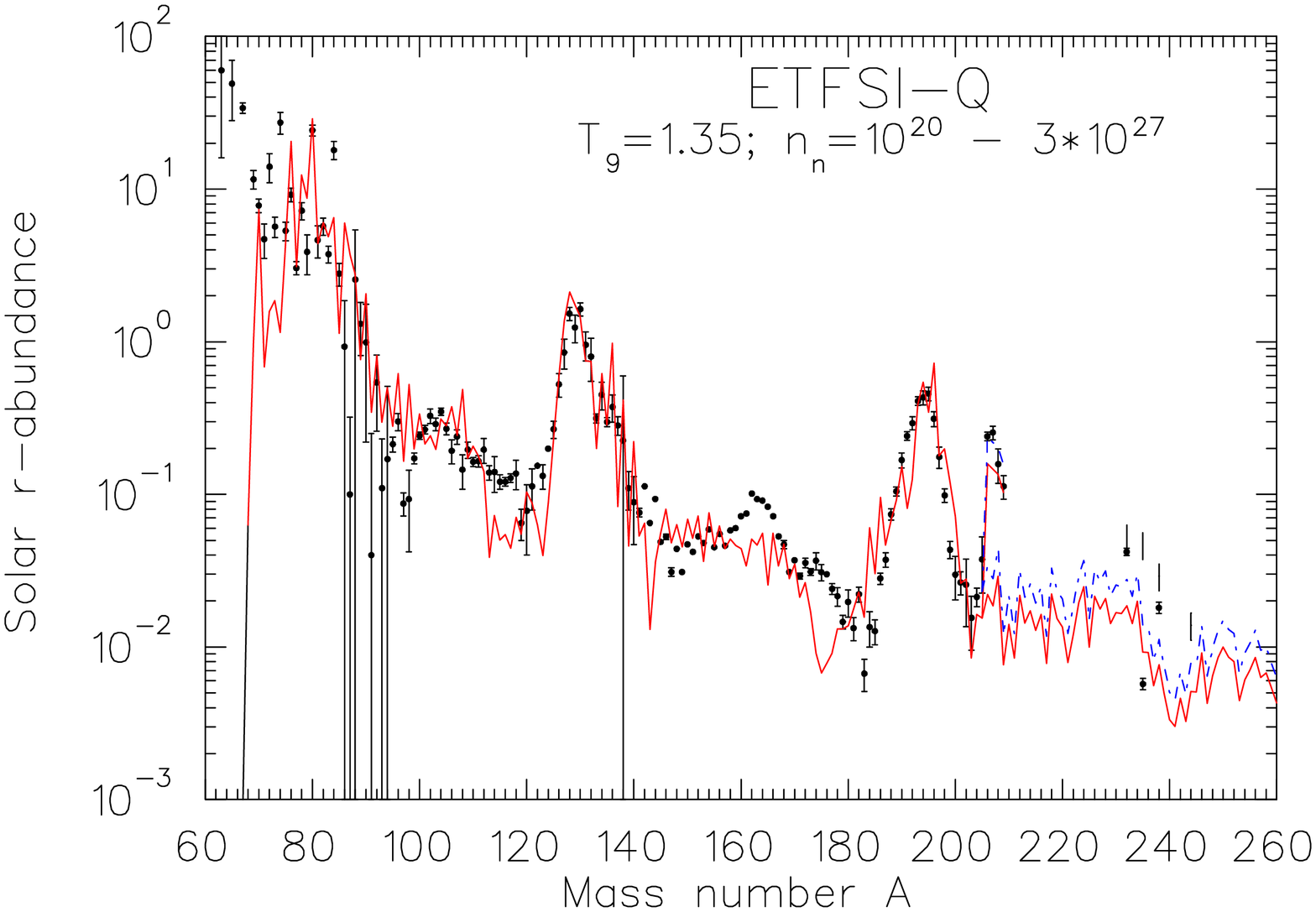,angle=90,width=15cm}}
\caption[F2]{Fits to solar r-process abundances,
obtained with two different smooth superposition of 17 equidistant 
$S_n(n_n,T)$ components from 1 to 4 MeV (solid and dashed lines).
The ETFSI-Q mass model~\cite{pearson96} was applied,
which introduces a phenomenological quenching of shell effects.
The quenching of the $N=82$ shell gap avoids a large abundance
trough below the $A$=130 peak. These results also
show a good fit to the r-process
Pb and Bi contributions after following the decay chains of unstable heavier
nuclei (indicated by two sets of abundances for A$>$205).
\label{ueberlagerung}}
\end{figure}
A similar effect was found recently in further self-consistent burst 
calculations~\cite{fisker00,rembges00}, when REACLIB rates in the
mass range $A$=44--63 were
replaced by cross sections based on predictions of resonance properties
from shell model calculations~\cite{fisker00a}. These effects are even more
drastic, again due to the early burning phase when matter beyond Ne
burns up to Fe, before the alpha-captures begin. This shows that
a more precise determination of specific reaction rates is important, when
self-consistent network plus hydrodynamics calculations are performed.
The peak temperatures and densities attained in X-ray burst calculations
depend on this cross sections input.

\begin{figure}[th]
\centerline{\epsfig{file=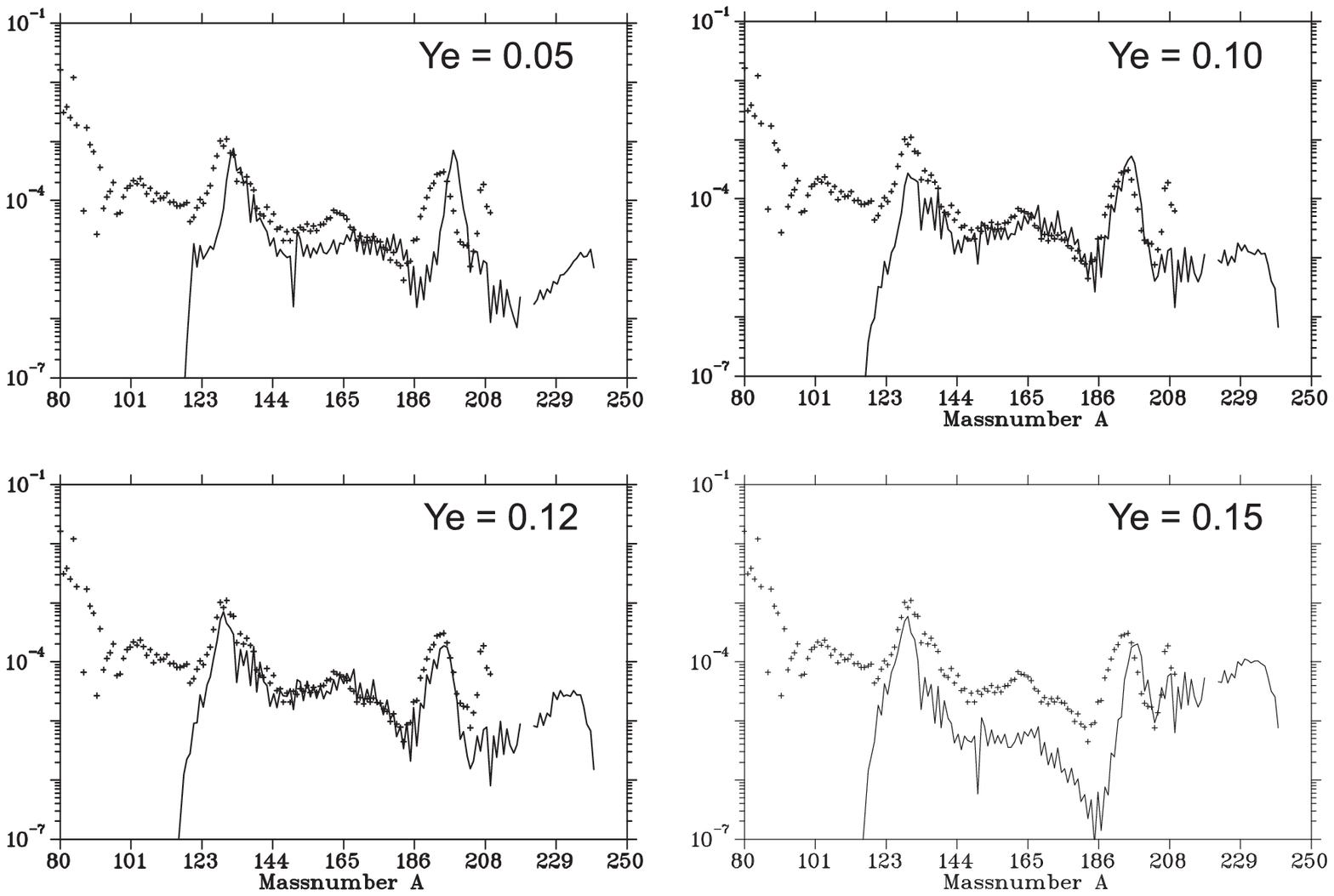,angle=0,width=16cm}}
\caption[F2]{Calculated r-process distribution for
different $Y_e$'s. In general one obtains useful contributions for
$0.08<Y_e<0.15$. A further discussion is given in the text.
$Y_e$ determines the total neutron/seed ratio, which is an indication
of the strength of the r-process. It affects also the combination
of $n_n$ and $T$, i.e.\ the r-process path, and therefore the position
of peaks. Finally, fission cycling is responsible for the drop of abundances
below A=130, but only an improved incorporation of fission barriers and 
yields will provide the correct abundance distribution in this mass range.}
\label{fig4}
\end{figure}
If only a small percentage of the synthesized matter escapes the strong
gravitational field of the neutron star, some proton-rich stable nuclei
(p-process nuclei) below $A$=100 (indicated
as P in Fig.\ref{rp1}) could be explained in the solar system abundances.

\section{The r-Process}

\subsection{Abundances and Nuclear Properties far from Stability}
Site-independent classical analyses,  
based on neutron number density $n_n$, temperature $T$, and duration time
$\tau$, led to the conclusion that the r-process experienced a fast drop from 
an $(n,\gamma)-(\gamma,n)$ chemical equilibrium in each isotopic chain.
The combination of $n_n$ and $T$ determines an r-process path
in the nuclear chart along nuclei with a neutron separation energy $S_n(n_n,T)$.
Thus, the r-process and its abundance features probe nuclear structure
far from stability via mass properties and the beta-decay half-lives along
contour lines of constant $S_n$ \cite{kratzea93}. This gives some indication
for the need of quenching of nuclear shell effects far from stability
\cite{pfeiffer97,kratz00,kratz01}.
A continuous superposition of components with neutron separation 
energies in the range 1--4~MeV on 
timescales of 1--2.5 s, provides a good overall fit \cite{cowan99}.
For the heavier elements beyond A=130 this is reduced to about  $S_n$=1--3 MeV.
These are predominantly nuclei not accessible in 
laboratory experiments to date. Exceptions exist in the $A=80$ and 130 peaks
and continuous efforts
are underway to extend experimental information in these regions of the closed
shells $N$=50 and 82 with radioactive ion beam facilities \cite{kratz00}.
A recent detailed analysis of the A=206--209 abundance contributions
to Pb and Bi isotopes from alpha-decay chains of heavier nuclei permitted 
for the first time also to predict abundances of nuclei as heavy as Th
with reasonable accuracy \cite{cowan99}. These results are shown in Fig.~7.

The endpoint of the r-process path is determined by $\beta$-delayed
fission when the path reaches fissionable nuclei. Fission fragments may
alter not only the heavy element abundances but may also give rise to an
exponential enhancement of the general r-process abundances, in sites
where an extended duration of the high-flux neutron supply leads to
fission cycling, such as in some inhomogeneous big bang models \cite{rauscher94}
or neutron star mergers (see the next section). The knowledge of
fission barriers and an improved theoretical treatment of
$\beta$-delayed fission (involving fission of highly excited nuclei) is
crucial for the future investigation of these effects.

\subsection{Possible Stellar r-Process Sites}
\label{pstudy}

A different question is related to the actual astrophysical realization
of such conditions.
The observations of stellar spectra of low metallicity stars,
stemming from the very early phases of galactic evolution,
are all consistent with a solar r-abundance pattern for elements heavier than 
Ba, and the relative abundances among heavy elements apparently do not show 
any time evolution \cite{cowan99,sneden00}. 
This suggests that all contributing events 
produce the same relative r-process abundances for the heavy masses, although 
a single astrophysical site will still have varying conditions in different 
ejected mass zones, leading to a superposition of individual components. 

However, from meteoritic abundances and observations in low metallicity stars
we also know by now that at least two r-process sources have to contribute to 
the solar r-process abundances \cite{wasserburg96,sneden00}.
The observed non-solar r-process pattern for e.g.\ Ag, I, and Pd in some 
objects indicate the need for a second r-process component in the nuclear mass
range A$\approx$80--120, in addition to the main process which 
provides a solar r-process pattern beyond Ba \cite{sneden00}. It is not 
exactly clear which
of the two processes is related to SNe II and which one is related to
possible other sources.

An r-process requires 10 to 150 neutrons per seed nucleus (in the Fe-peak or 
somewhat beyond) which have 
to be available to form all heavier r-process nuclei by neutron capture. 
For a composition of Fe-group nuclei and free neutrons that
translates into a $Y_e=$$<$$Z/A$$>$$=$0.12--0.3. Such a high neutron excess is
only possible for high densities in neutron stars under beta equilibrium
($e^-+p\leftrightarrow n+\nu$, $\mu_e+\mu_p=\mu_n$),
based on the high electron Fermi energies which are
comparable to the neutron-proton mass difference. 
Neutron star mergers which eject such matter are a possible (low entropy) site 
~\cite{freiburghaus99a}
and have been debated in the past. Recent calculations show that on average 
about $10^{-2}$ M$_\odot$ of neutron-rich matter
are ejected \cite{rosswog99,rosswog00}. This amount depends on the central 
high density equation of state \cite{prakash97} encountered in these events.
Present calculations show densities up to
four times nuclear matter density and temperatures of up to 50~MeV 
\cite{rosswog99,rosswog00}.
First nucleosynthesis calculations with assumptions on $Y_e$ 
predict a solar-type r-process pattern for
nuclei beyond A=130 \cite{freiburghaus99b}, shown in Fig.~8. The smaller 
masses are depleted due to a long duration
r-process with a large neutron supply in such neutron-rich matter, which
also leads to fission cycling~\cite{panov01}. This seems (accidentally?) in 
accordance with the main observed r-process component.
Given the frequency ($10^{-5}$y$^{-1}$ per galaxy)
and amount of ejected matter, this component alone
could be responsible for the heavy solar r-process pattern and also
explain the large scatter of r/Fe elements found in low metallicity stars
\cite{truran00}.
Neutron star - black hole mergers have not yet been analyzed with the
same accuracy, but bear similar options.

Another option is an extremely alpha-rich (i.e.\ high entropy) freeze-out in 
complete Si-burning with moderate $Y_e$$>$0.40, which however faces some of
the problems already mentioned in the section on SNe II
\cite{freiburghaus99a,mclaughlin99}.
A discussion of the advantages and disadvantages of both possible r-process
sources (SNe II vs.\ neutron star mergers) is given in refs. 
\cite{qian00,rosswog01}.

\section{Conclusions}

This overview concentrated on nuclear physics issues important in stellar
evolution, supernovae (type II and Ia), X-ray bursts and the rp-process,
and analyzed the options and sites of r-process
nucleosynthesis. These are the major contributions to galactic evolution.
Nucleosynthesis calculations have a right on their own to predict abundance
patterns for many stellar events, but they can also serve as a tool to test
the correctness of model descriptions, either in comparison to direct 
observations or indirect information from galactic evolution. We tried to show
(especially for SNe Ia and II), how specific isotopic abundances can test
ignition densities, burning front velocities or explosion energies, entropies,
and temperatures. These are the astrophysical model constraints. But it was
also clearly demonstrated how advances in nuclear physics (nuclear
reaction cross sections, weak interaction rates in the pf-shell, 
neutrino-nucleus interactions, decay properties far from stability, nuclear
structure far from stability, fission properties and yields, the nuclear
equation of state at high densities and temperatures) are essential for
the outcome and correct modeling of these events.

This review would not have been possible without discussing and presenting
results from joint collaborations with J.J. Cowan, D. Dean, 
R.D. Hoffman, K. Iwamoto, F. K\"appeler, M. Strayer, J.W. Truran, and
S.E. Woosley.

\end{document}